\documentstyle[11pt,aaspp4,psfig]{article}

\def \ha {H$\alpha$}
\def \hb {H$\beta$}

\def \oiii  {{\rm [O~III]}}
\def \oii   {{\rm [O~II]}}
\def \oi    {{\rm [O~I]}}
\def \nii   {{\rm [N~II]}}
\def \sii   {{\rm [S~II]}}
\def\wave#1{$\lambda${#1}}
\def\waves#1{$\lambda\lambda${#1}}
\def\etal{{\rm et al.}}

\def\arcsecpoint{\hbox to 1pt{}\rlap{\arcsec}.\hbox to 2pt{}}
\def\ltsima{$\; \buildrel < \over \sim \;$}
\def\simlt{\lower.5ex\hbox{\ltsima}}
\def\gtsima{$\; \buildrel > \over \sim \;$}
\def\simgt{\lower.5ex\hbox{\gtsima}}

\def   \flam    {$F_\lambda$}
\def   \pf      {$P\times F_\lambda$}
\def   \qf      {$Q\times F_\lambda$}
\def   \uf      {$U\times F_\lambda$}
\def   \kms     {km s$^{-1}$}
\def   \p       {$P$}

\def   \pa      {$\theta$}

\def \fluxu {ergs cm$^{-2}$ s$^{-1}$ \AA$^{-1}$} 
\def \lsun      {$L_{\sun}$}
\def \msun      {$M_{\sun}$}
\def \firr      {$f_{25}/f_{60}$}
\def \hubu      {km s$^{-1}$ Mpc$^{-1}$}

\slugcomment{Accepted for publication in the {\it Astrophysical Journal}}

\begin{document}

\title{A Polarimetric Search for Hidden Quasars in Three Radio Selected 
Ultraluminous Infrared Galaxies}
\author{Hien D. Tran\altaffilmark{1}, 
M. S. Brotherton\altaffilmark{1}, 
S. A. Stanford\altaffilmark{1},
Wil van Breugel\altaffilmark{1},
Arjun Dey\altaffilmark{2}, 
Daniel Stern\altaffilmark{3},
Robert Antonucci\altaffilmark{4} 
}
\altaffiltext{1}{Institute of Geophysics and Planetary Physics, Lawrence 
Livermore National Laboratory, 7000 East Avenue, P.O. Box
808, L413, Livermore, CA 94550; htran@igpp.llnl.gov.} 
\altaffiltext{2}{National Optical Astronomy Observatories, P.O. Box 26732, 
Tucson, AZ 85726-6732.}
\altaffiltext{3}{Department of Astronomy, University of California, Berkeley, 
CA 94720-3411.} 
\altaffiltext{4}{Department of Physics, University of California, Santa Barbara,CA 93106.} 


\begin{abstract}

We have carried out a spectropolarimetric search for hidden broad-line quasars
in three ultraluminous infrared galaxies (ULIRGs) discovered in the positional
correlations between sources detected in deep radio surveys and the 
$IRAS~Faint~Source~Catalog$. 
Only the high-ionization Seyfert 2 galaxy TF~J1736+1122 is highly polarized, 
displaying a broad-line spectrum visible in polarized light. 
The other two objects,
TF~J1020+6436 and FF~J1614+3234, display spectra dominated by a population of 
young (A-type) stars similar to those of ``E + A'' galaxies. 
They are unpolarized, showing no sign of hidden broad-line regions.
The presence of young starburst components in all three galaxies indicates 
that the ULIRG phenomenon encompasses both AGN and starburst 
activity, but the most energetic ULIRGs do not necessarily harbor 
``buried quasars''. 

We find that a luminous infrared galaxy is most likely to host an 
obscured quasar if it exhibits a high-ionization 
(\oiii \wave 5007/\hb~\simgt~5) spectrum 
typical of a ``classic'' Seyfert 2 galaxy with little or no Balmer absorption 
lines, is ``ultraluminous'' ($L_{IR} \simgt 10^{12}$ \lsun), and has a
``warm'' IR color ($f_{25}/f_{60}$~\simgt~0.25). 
The detection of hidden quasars in this group but not in the low-ionization,
starburst-dominated ULIRGs (classified as LINERs or H II galaxies)
may indicate an evolutionary connection, with the latter being found in
younger systems.

\end{abstract}


\section{Introduction}

Many sources in the $IRAS~Faint~Source~Catalog$ ($FSC$) exhibit enormous
infrared luminosities ($L_{IR} > 10^{12}$ \lsun), and are known as
ultraluminous infrared galaxies (ULIRGs, \cite{san88}).  These objects
are an important constituent of our local Universe, with luminosities
and space densities similar to those of quasi-stellar objects (\cite{soi87}). 
This led to the suggestion that ULIRGs could contain
infant quasars enshrouded in a large amount of dust (\cite{san88}). On
the other hand, they may also represent energetic, compact starbursts
(\cite{con91}).  Understanding the dominant energy input mechanism in
these ULIRGs -- whether it is obscured quasars or intense bursts of
star formation --  has remained the main issue concerning their nature.
Of particular interest are the ``warm'' ULIRGs (i.e., those having
$f_{25}/f_{60}~>~0.2$, \cite{low88}; \cite{san88})\footnote{$f_{25}$ and $f_{60}$ are the $IRAS$ flux densities in units of Jy at 12 $\mu$m
and 60 $\mu$m, respectively.}.  They include a few of the exceptionally
luminous sources, sometimes referred to as the ``hyperluminous''
($L_{IR}~\simgt~10^{13}$~\lsun) infrared galaxies, such as P09104+4109
($z$ = 0.44; \cite{k88}), F15307+3252 ($z$ = 0.926; \cite{c94}), and
F10214+4724 ($z$ = 2.286; \cite{rr91}).  All of these galaxies have been
shown to harbor a quasar nucleus obscured from direct view through the
detection of hidden broad line regions (HBLRs) in polarized, scattered
light (P09104+4109: \cite{hw93}; Tran 1998; F15307+3252: \cite{hin95};
F10214+4724: \cite{goo96}).  This has led to the suggestion that perhaps
all warm ULIRGs contain buried QSOs and that they may be the misdirected
type 2 QSOs (AGN with Seyfert 2-like emission-line characteristics but
QSO-like luminosities; \cite{wh98}).

With the exception of the above extreme $IRAS$ sources, most of the known
ULIRGs are relatively nearby ($z$~\simlt 0.1).  In order to identify more
ULIRGs at higher redshifts and study their population, several searches
have been undertaken, using the positional correlation of sources in the
$FSC$ with those detected in deep radio surveys.  Stanford \etal~(1998)
correlated objects in the $FSC$ with those in the Faint Images of
the Radio Sky at Twenty-cm (FIRST) catalog (Becker, White, \& Helfand
1995), which has a flux density limit of $\approx$ 1 mJy. 
The results show that nearly all matches that are faint on the
Palomar Observatory Sky Survey (POSS) turn out to be ULIRGs, if they lie on the
well-known radio-to-far infrared (FIR) flux correlation (\cite{con91})
established by normal late-type galaxies and starburst galaxies at low
redshifts. Spectroscopically, the vast majority of the sources ($\sim$ 85\%)
in this ``radio-quiet'' FIR sample (hereafter FF sources) can be classified 
as starburst systems (\cite{sta98}).

Dey \& van Breugel (1994) correlated the Texas 365 MHz radio catalog 
(\cite{dou96}) with an early version of the $FSC$, and selected from this sample
those sources that lie {\it above} the radio-to-FIR flux-flux
correlation (see \cite{sa91}; \cite{cp95}; \cite{sta98}).  
From this sample (hereafter TF sources) they
discovered a significant class of gas and dust-rich AGN which are
characterized by large FIR luminosities, and which are intermediate
in radio luminosity, FIR color, and optical spectral class between the
$IRAS$ ultraluminous galaxies and powerful quasars (\cite{dvb94}; 
\cite{sta98}). Because of the much higher detection threshold of the Texas
survey ($\approx$ 150 mJy, \cite{dou96}), the majority of these sources are
``radio-loud'' LINERs, Seyfert 2s and post--starburst active galactic nuclei
(PSAGN).  The latter are characterized by the strong Balmer absorption
lines, Seyfert 2-like emission lines and compact, often almost stellar,
optical morphologies.

A notable feature of both the radio-quiet and radio-loud FIR starburst
and PSAGN objects, visible in high signal-to-noise spectra, is the
prevalence of many strong high-$n$ Balmer absorption lines (EW $\sim$
10~\AA), as well as a strong Balmer discontinuity in their spectra.
These features signify a large population of
hot, relatively young stars (10$^8$ -- 10$^9$ yrs).  In addition,
their emission-line spectra tend to exhibit a very low ionization
state, as characterized by the strength of \oii~\wave 3727 relative to
\oiii~\wave 5007 (\oii~\wave 3727/\oiii~\wave 5007 \simgt 2), and the
lack or weakness of high-ionization lines such as [Ne~V]~\wave 3426.
Qualitatively, the spectra of these galaxies appear very similar to those
of the class of galaxies called ``E + A'' or ``post-starburst'' galaxies
(\cite{dg83}; Oegerle, Hill, \& Hoessel 1991; Wirth, Koo, \& Kron 1994;
\cite{lg96}), whose spectra display not only the
4000~\AA~break, G-band, and MgI b features that are indicative
of an old ($\sim$ 10 Gyr) stellar population (the ``E'' component), but
also the Balmer absorption lines characteristic of much younger stars
($<$~1 Gyr, the ``A'' component).  Many FF and TF sources appear to differ
from E + A galaxies only in the additional presence of emission lines, 
indicating the presence of on-going star formation or active galactic nuclei 
(see spectra of TF~J1020+6436 and FF~J1614+3234 below). These objects may
form an important evolutionary link between starbursts and quasar activity
(\cite{cs97}; Stockton, Canalizo, \& Close 1998).

In order to better understand the energy sources of ULIRGs and the
relationship between AGN and starburst activity, we started a 
spectropolarimetric survey to search for hidden broad emission lines
in both types of objects: the radio-quiet (FF) and radio-loud (TF) ULIRGs.
This would indicate whether obscured AGNs might exist, and the high
signal-to-noise data might be used to investigate the nature of their stellar
populations. This might be used to help quantify the relative importance
of quasar and starburst activity, and to search for possible correlations
with the age of the starbursts, luminosity, and morphological evolution.

This paper reports the first results for three objects
with $L_{IR} \simgt 10^{12}$ \lsun\footnote{$L_{IR}$ is calculated by following
the prescription of Sanders \& Mirabel (1996).}: TF~J1020+6436, TF~J1736+1122,
and FF~J1614+3234. Only TF~J1736+1122 is certainly IR ``warm'' with
\firr~= 0.404. For TF~J1020+6436 and FF~J1614+3234, only an upper limit
of the $f_{25}$ flux density was obtained.
Table 1 lists the infrared properties for the three objects.
Throughout this paper, we assume $H_o$ = 75 \hubu, $q_o$ = 0 and $\Lambda$ = 0.

\section{Observations and Reductions}

\subsection{Spectropolarimetry}

Spectropolarimetric observations were made with the polarimeter 
(\cite{coh97}) installed in the Low Resolution Imaging Spectrometer 
(LRIS, \cite{oke95}) on the 10-m Keck II telescope on the night 
of 1997 April 10 (UT). A 1\arcsec~wide, long slit oriented at the parralactic
angle was centered on the nucleus of each galaxy. We used a 300 grooves 
mm$^{-1}$ grating which provided a dispersion of 2.46~\AA~pixel$^{-1}$ 
and a resolution of $\sim$~10~\AA~(FWHM).
The observations were made by following standard procedures of rotating 
the half waveplate to four position angles (0\arcdeg, 22.5\arcdeg, 45\arcdeg, 
and 67.5\arcdeg), and dividing the exposure times equally among them.  
The total exposure times were one hour each for TF~J1020+6436 and
FF~J1614+3234, and 40 minutes for TF~J1736+1122.  
We followed standard polarimetric reduction techniques as described, 
for example, by Cohen \etal~(1997).

\subsection{Near-Infrared Imaging}

Near-IR observations of FF~1614+3234 were made using the Near Infrared Camera 
(NIRC, \cite{ms94}) at the Keck I Telescope. NIRC employs a 256$\times$256 
InSb array with a scale of 0.15~arcsec~pixel$^{-1}$.  The data were
obtained through a $K_s$ filter on 1998 April 18 (UT), under photometric
conditions with $\sim$0\farcs5 seeing.  The target exposures were taken
using a nonredundant dithering pattern, with individual shifts between
pointings of 3\arcsec.  Integration times for each pointing were
typically 60~s comprised of 3 coadds; the total on-source exposure
time was 1920~s.  After bias subtraction, linearization, and
flatfielding (using sky flats), the data were sky--subtracted,
registered and summed using the DIMSUM\footnote{DIMSUM is the Deep Infrared 
Mosaicing Software package, developed by P.\ Eisenhardt, M.\ Dickinson, 
A.\ Stanford, and J.\ Ward, and is available as a contributed package 
in IRAF.} near--IR data reduction package.  Subsequent photometry was flux 
calibrated via short observations of UKIRT faint standards (\cite{ch92}), 
which, after a suitable transformation, yields magnitudes on the CIT system.

\begin{figure}
\plotone{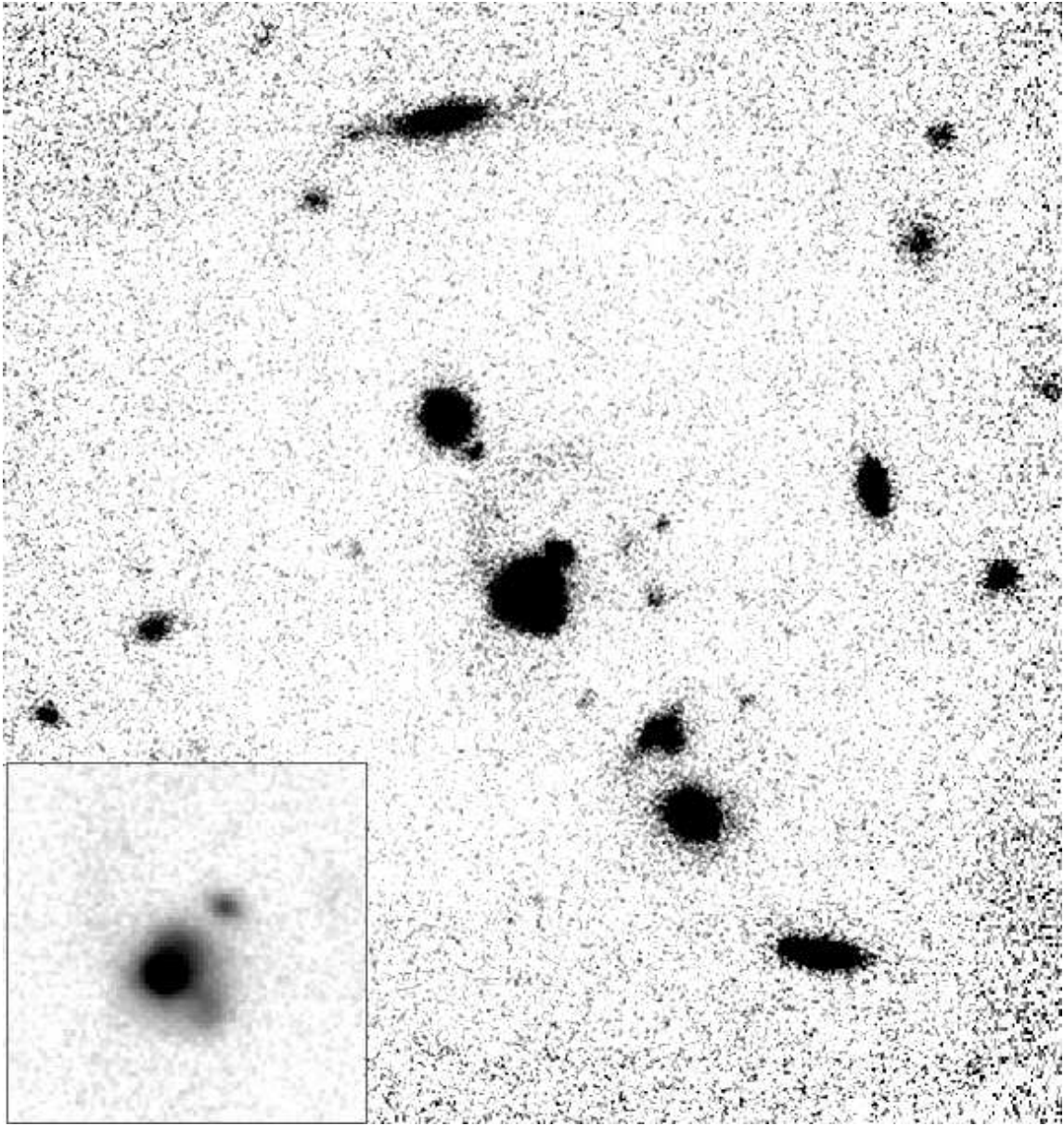}
\caption{K$_s$-band image of the field of FF~J1614+3234 obtained with the 
W. M. Keck I Telescope. North is up and east is to the left. 
FF~J1614+3234 is centered on the image. The inset displays a 
10\arcsec$\times$ 10\arcsec~region centered on
FF~J1614+3234 at a lower stretch, showing emission extending 1\arcsecpoint8~to
the south-west. Also apparent are a companion 2\arcsecpoint2~to the northwest 
and low-level filamentary extensions to the north. \label{ff16kb}}
\end{figure}

Figure \ref{ff16kb} shows  
the $K_s$-band image of FF~J1614+3234 obtained with Keck I Telescope. 
As can be seen, the near-IR image shows a disturbed morphology,
with an extension separated $\sim$ 1\arcsecpoint8~from the nucleus to the
south-west. At a lower surface brightness level, extended filamentary arc-like structure 
is present to the north and west. This feature is suggestive of a tidal tail 
due to a merger event or interaction.  
There is also an apparent companion $\sim$ 2\arcsecpoint2~to the north-west. 
These features are
consistent with FF~J1614+3234 being similar to an E + A galaxy, which 
often occurs in merging systems (e.g., \cite{lk95}). They
also provide strong circumstantial evidence that any starburst and AGN 
activity present is triggered by galaxy interaction.
No indication of gravitational lensing is evident from the image. 
 
\subsection{Radio Observations}

\begin{figure}
\plotone{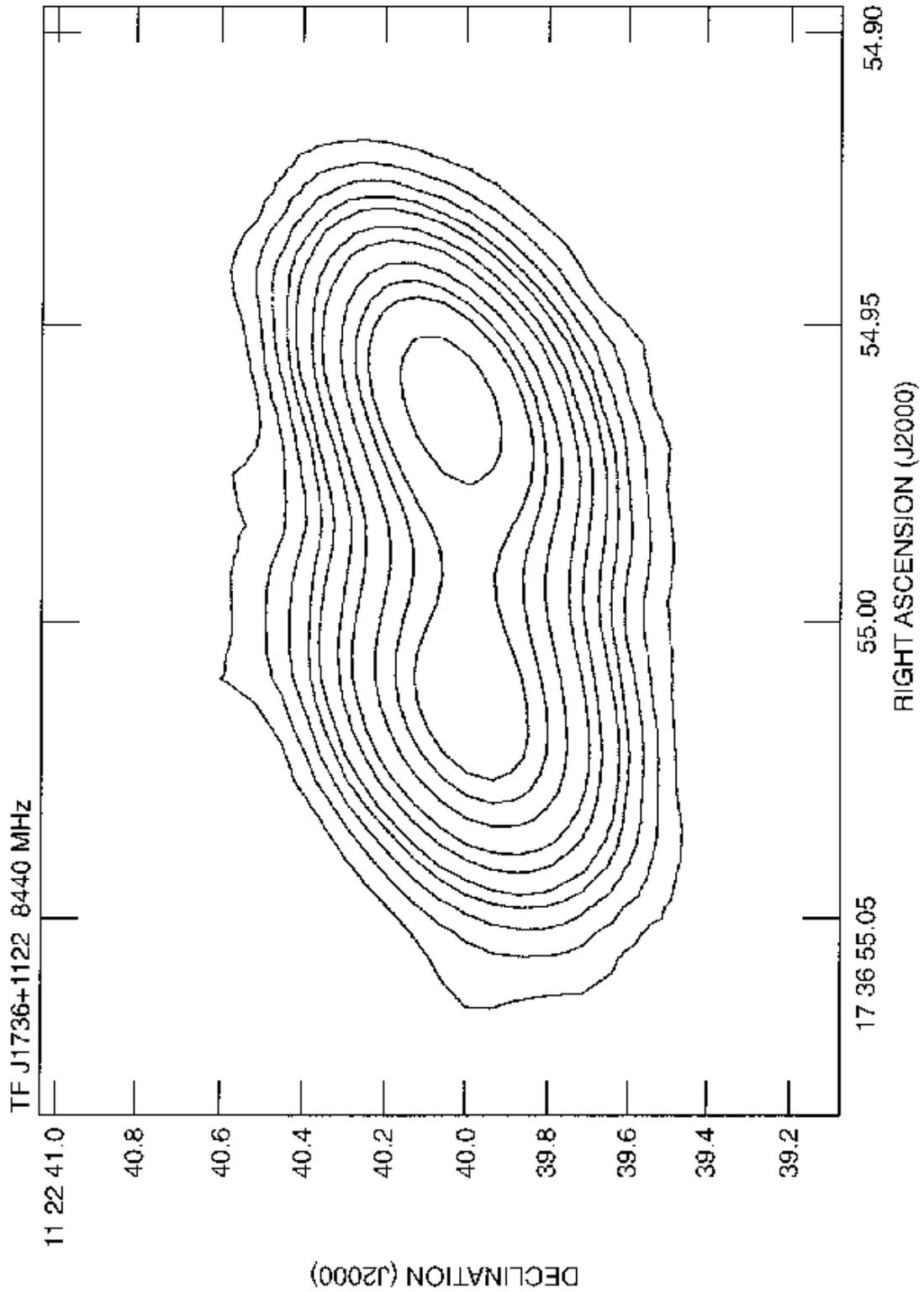}
\caption{The 8440 MHz VLA map of TF~J1736+1122. The contour levels are:
0.04 $\times$~(3, 6, 10, 15, 20, 30, 40, 60, 80, 100, 150) mJy beam$^{-1}$. The peak flux is at 7.77$\times$10$^{-3}$ Jy. \label{tf1734radmap}}
\end{figure}

TF~J1020+6436 and TF~J1736+1122 were both observed with the VLA, using
standard bandwidths and calibrations, to obtain more accurate positions,
and morphological and spectral index information.  TF~J1020+6436 was
observed at 4860~MHz in the C-array and was found to be unresolved with
an angular size $<$~1\arcsecpoint1. TF~J1736+1122 was observed at 4860~MHz
and 8440~MHz in the B and A-arrays respectively. The B-array observations
showed that this source was slightly resolved, with an angular size of
0\arcsecpoint9~$\times$ 0\arcsecpoint4, at PA~=~89\arcdeg.  
In the A-array, and at the higher frequency of 8440~MHz, the source shows 
a resolved core with a clear double structure (see Fig. \ref{tf1734radmap}) 
elongated along PA~$\sim$~97\arcdeg. This structure axis is essentially 
perpendicular ($\Delta$PA $\approx$ 80\arcdeg) to the polarization 
PA of 18\arcdeg~(see \S 4.1), as has been found for many polarized narrow-line 
radio galaxies and Seyfert 2 galaxies (e.g., \cite{t98} and references therein).
For FF~J1614+3234 the radio data was taken 
from the FIRST catalog (Becker \etal~1995). Only the 1400 MHz observations 
are available for this source, which is unresolved with an angular size of 
$<$ 4\arcsec. The positions and flux densities of the three sources are listed 
in Table 2.

It is of interest to note that the radio continua of the two radio-loud
TF sources have quite steep spectra, with spectral indices $\alpha \sim -1$
($f_\nu \propto \nu^\alpha$). In the case of TF~J1736+1122 the radio
spectrum steepens significantly between 365 MHz and 8440 MHz, with
$\alpha$ changing from $-$0.75 to $-$1.45 (see Table 2). In addition, both 
objects are quite small ($<$ 2.6 kpc). Such compact 
steep spectrum (CSS) radio sources comprise a significant fraction of 
complete radio source surveys, and are thought to be ``frustrated'' or 
young radio sources whose jets are plowing through the dense ISM of their 
parent galaxies (see O'Dea 1998 for a recent discussion). 

\section{Results}

\subsection{Line Ratios, Diagnostics, and Classification}

Table 3 presents the integrated emission-line flux ratios and their 
rest-wavelength equivalent widths. 
The emission-line fluxes were measured from the starlight-subtracted spectra.
For FF~J1614+3234 and TF~J1020+6436, isochrone population synthesis 
models of the underlying young starbursts (see \S 3.2) have been removed. 
For TF~J1736+1122, the contribution from an old 
stellar population represented by the elliptical galaxy NGC 4478 was 
used (\S 3.3). In TF~J1736+1122, the narrow emission lines display 
extended wings that cannot be adequately fit with a single Gaussian profile. 
A sum of two Gaussian profiles, one with FWHM $\sim$ 500 \kms~and 
another with FWHM $\sim$ 1300 -- 1800 \kms~but identical center, 
is able to reproduce the observed profiles very well.
The observed spectrum of TF~J1020+6436 appears to suffer significant internal 
reddening. The measured narrow-line Balmer decrement is \ha/\hb~= 8.8,
giving an extinction of $E(B-V) = 1.05$, assuming an intrinsic ratio 
of \ha/\hb~= 3.1 and using the extinction law of Cardelli, Clayton, \& 
Mathis (1989).

Plotting the line flux ratios 
(involving \oii~\wave 3727, \hb, \oiii~\wave 5007, 
\nii~\waves 6548, 6583, \ha, \sii~\wave 6724, and \oi~\wave 6300)
on the diagnostics diagrams of Baldwin, Phillips, \& Terlevich (1981) and Veilleux \& Osterbrock (1987) 
shows that both TF~J1020+6436 and TF~J1736+1122 lie in regions typically 
occupied by narrow-line AGNs. 
This suggests that the dominant energy input in the narrow line regions (NLRs)
is photoionization by a hard continuum. 
These objects can be classified as Seyfert 2 galaxies.
However, only TF~J1736+1122 can be called a {\it bona fide} 
high-ionization Seyfert 2. Although TF~J1020+6436 formally qualifies as a 
type 2 Seyfert, its ionization level is quite low
(\oii~\wave 3727/\oiii~\wave 5007~= 1.4).
What clearly separates it from TF~J1736+1122 is the domination of 
early-type stellar spectral features in the blue end of the spectra.
TF~J1020+6436 satisfies half of the definition of Heckman (1980) 
for LINERs\footnote{Criteria for LINERs as defined by Heckman (1980) are:
\oii~\wave 3727/\oiii~\wave 5007~$>$~1 and 
\oi~\wave 6300/\oiii~\wave 5007~$>$~1/3.}: \oii/\oiii~$>$~1. 
The other half of the criterion is not satisfied 
(\oi/\oiii~=~0.15~$<$~0.33, Table 3), and thus it is not a LINER, and we
adopt the Seyfert 2 classification for it. 
Since our spectrum of FF~J1614+3234 does not cover wavelengths longward of
\oiii~\wave 5007, an unambiguous classification of this galaxy based on the 
line ratios of \nii, \ha, and \sii~is not possible.
Using the available line strengths, we classify FF~J1614+3234 as a LINER based
on the low ionization level (\oii~\wave 3727/\oiii~\wave 5007~=~1.75; 
\oiii~\wave 5007/\hb~=~2.0) and the strong appearance of hot-star spectral 
features.

\subsection{No Detection of Hidden Quasars in FF~J1614+3234 and TF~J1020+6436}

\begin{figure}
\plottwo{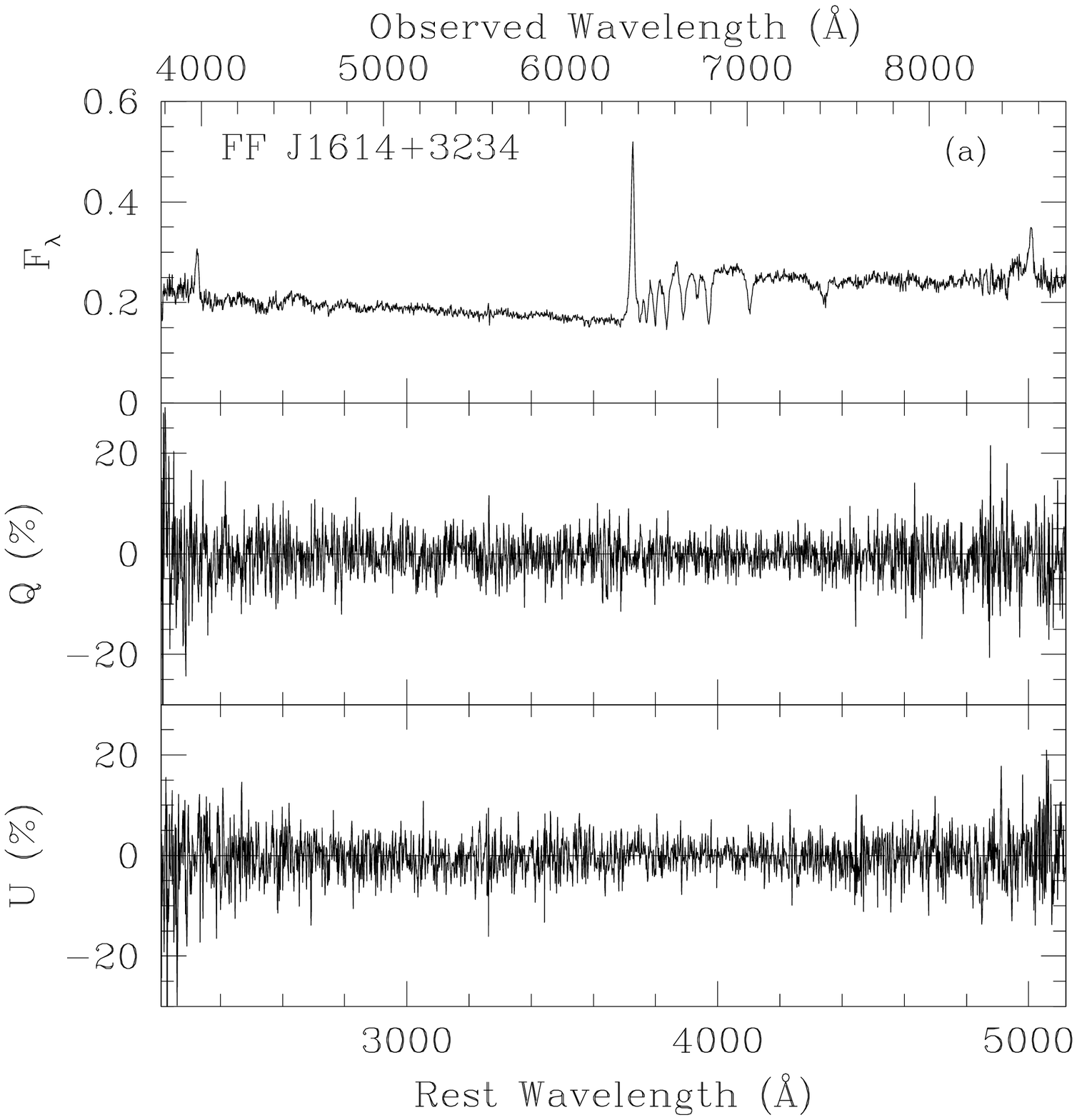}{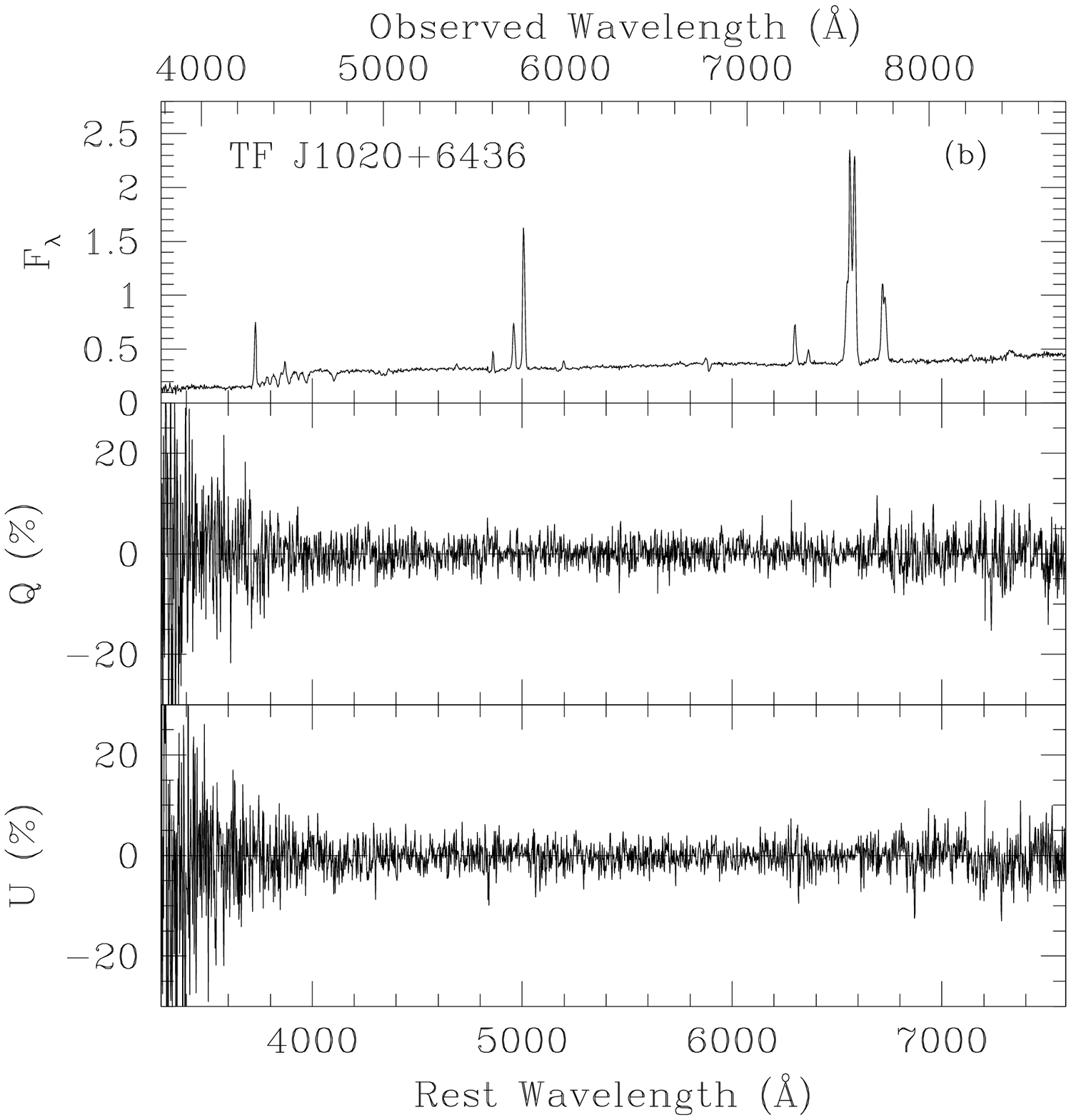}
\caption{Optical spectra and normalized $Q$ and $U$ Stokes parameters for 
(a) FF~J1614+3234 and (b) TF~J1020+6436 obtained at the W. M. Keck II Telescope.
The flux scales are in units of 10$^{-16}$ \fluxu. 
Note the strong Balmer absorption lines characteristic of young stars in the 
spectra. No significant polarization is detected. \label{fftfspec}}
\end{figure}

FF~J1614+3234 and TF~J1020+6436, whose total flux spectra and $Q$ and $U$ 
Stokes parameters are shown in Figure \ref{fftfspec}, display no significant 
detectable polarizations. 
A formal averaging of the continuum polarization over the observed
wavelength range 4500~\AA~-- 8000~\AA~gives \p~= 0.24\% $\pm$ 0.18\%
and 0.15\% $\pm$ 0.15\% (uncorrected for biasing)
in FF~J1614+3234 and TF~J1020+6436, respectively.
As a result, we detected no polarized broad lines in the spectra of two of 
the more luminous ULIRGs in our sample.
There is also little sign of Mg II~\wave 2800~in the spectrum of 
FF~J1614+3234. It shows a very weak feature at the wavelength expected 
for Mg II~\wave 2800, but the profile is highly irregular.
Although the non-detection of hidden broad lines, by itself, does not
rule out the presence of a quasar, as the geometry may not be
propitious for scattered light to reach our line of sight (e.g., blockage
or lack of scattering region, high viewing inclination), 
the low excitation state and the presence
of strong Balmer absorption lines seem to indicate that the energy
output is dominated by a young starburst.
The enormous IR luminosity is likely the result of this starburst activity.

\begin{figure}
\plottwo{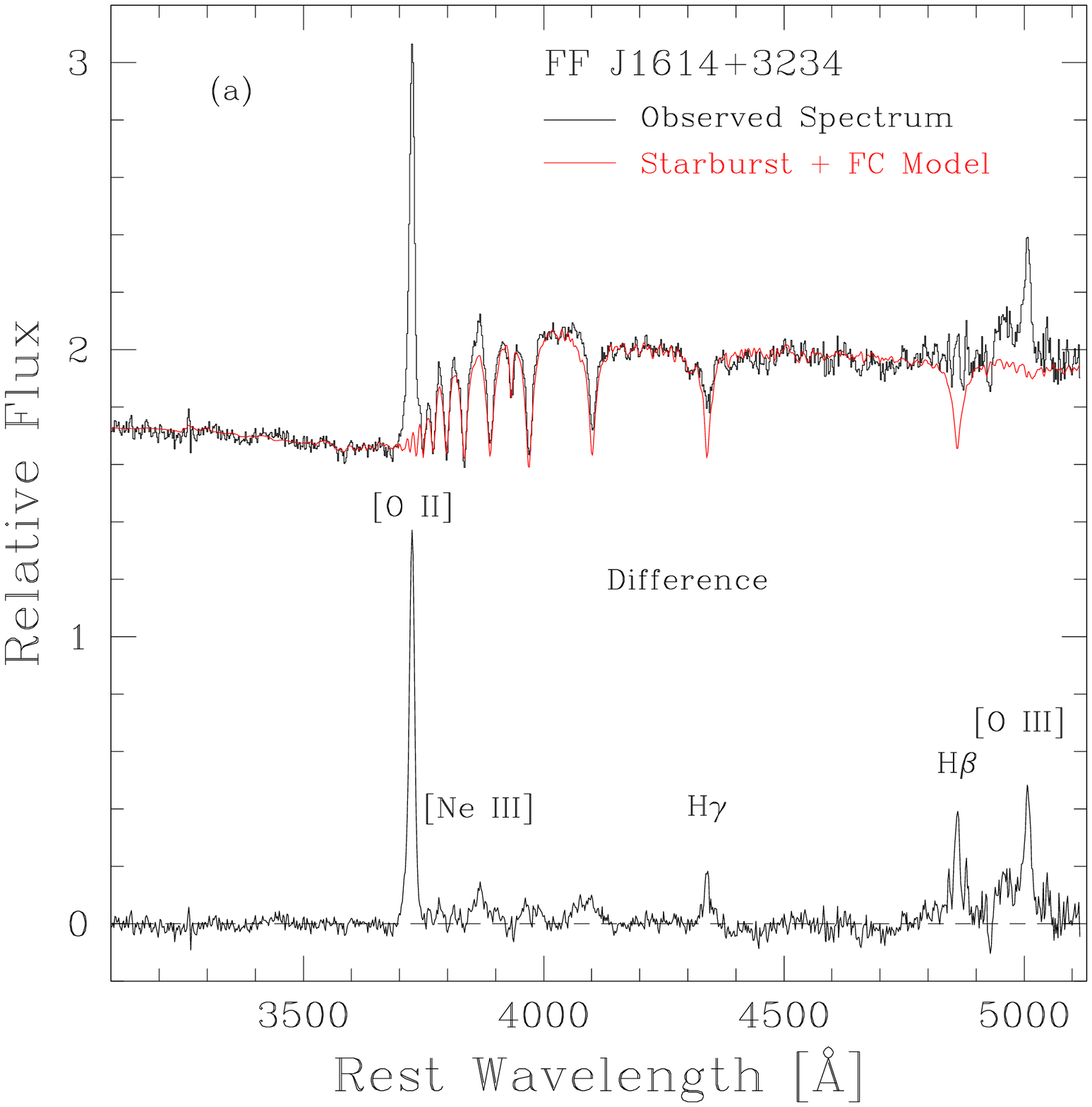}{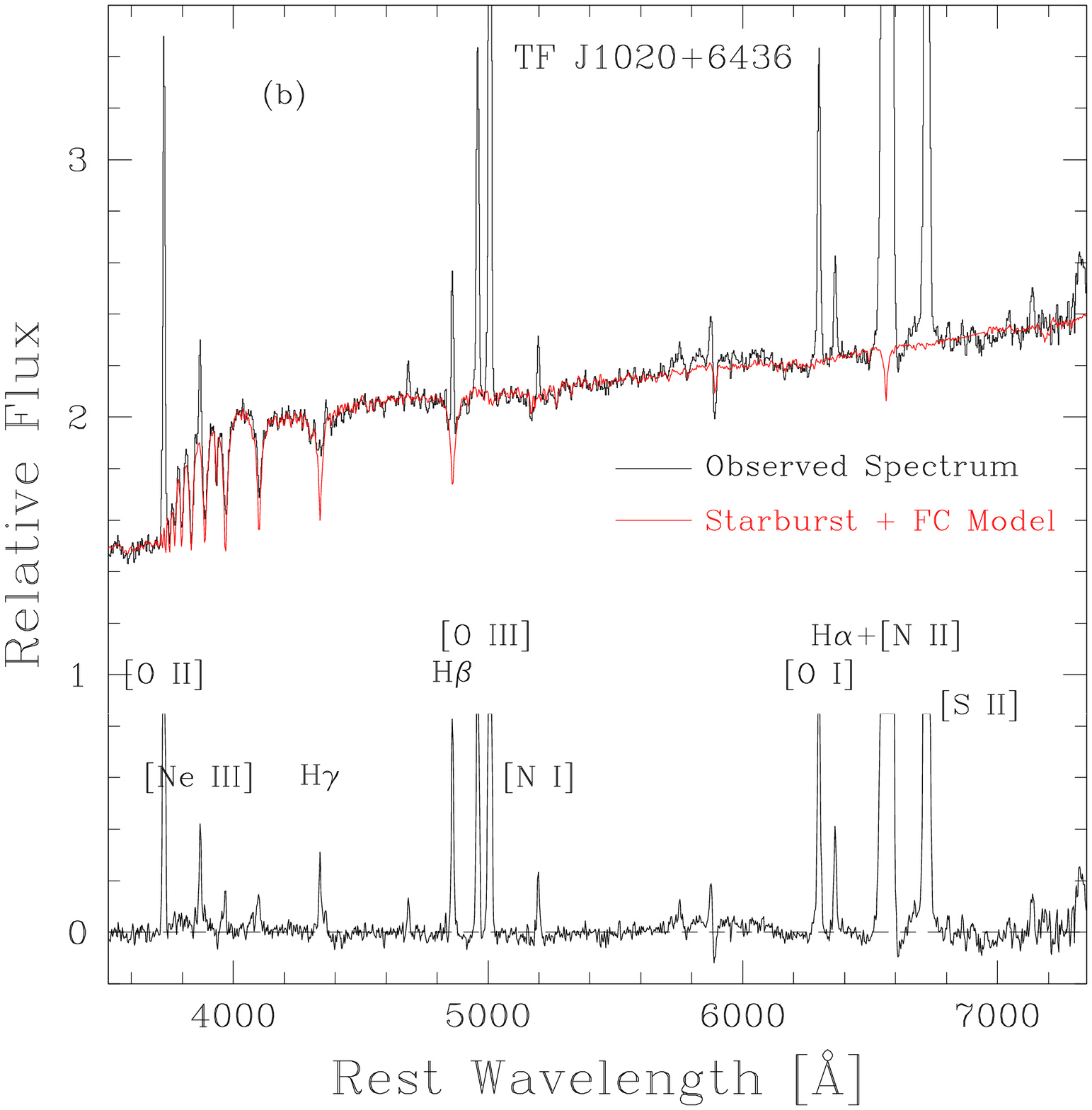}
\caption{Fits of BC96 models to the spectra of (a) FF~J1614+3234 and 
(b) TF~J1020+6436. Note the appearance of emission lines of 
[Ne III]~\wave 3869, H$\delta$, H$\gamma$, and \hb~after removal of the 
young starburst component. \label{bcmod}}
\end{figure}

We used the GISSEL models of Bruzual \& Charlot (1996, BC96) to 
characterize the young stellar population dominating the spectra of
FF~J1614+3234 and TF~J1020+6436. 
We assumed an instantaneous burst of star formation with a 
Salpeter (1955) initial mass function having a mass range 
$0.1 M_{\sun} < M <125 M_{\sun}$ and solar metallicity. 
The fit is a $\chi^2$ minimization of the sum of a blue featureless continuum 
(representing the Seyfert 2/LINER ionizing continuum) and the young starburst 
model. Our analysis indicates that some additional reddening is also 
required for a good fit.
Using the starburst extinction curve of Calzetti, Kinney, \& Storchi-Bergmann 
(1994), the required reddening is $E(B-V) \sim 0.3$ for FF~J1614+3234 and 
$\sim$ 0.35 for TF~J1020+6436. 
Figure \ref{bcmod} shows the best fits to the observed spectra, which  
were achieved for a starburst with an age of 320 Myr.  
This is consistent with the median age derived for the starburst knots found
in all of the ULIRGs studied by Surace \etal~(1998) using $HST$ imaging. 
Our results show that the light from the 320 Myr-old 
starburst component contributes about 68\% and 76\% of the total light 
at 4200~\AA~(rest frame) for FF~J1614+3234 and TF~J1020+6436, respectively. 
The rest of the light is due to the blue ionizing continuum of the 
Seyfert 2/LINER nucleus and perhaps an even younger population of hot (OB) 
stars.

\subsection{TF~J1736+1122: Hidden Broad Lines and Starburst Activity}

\begin{figure}
\plotone{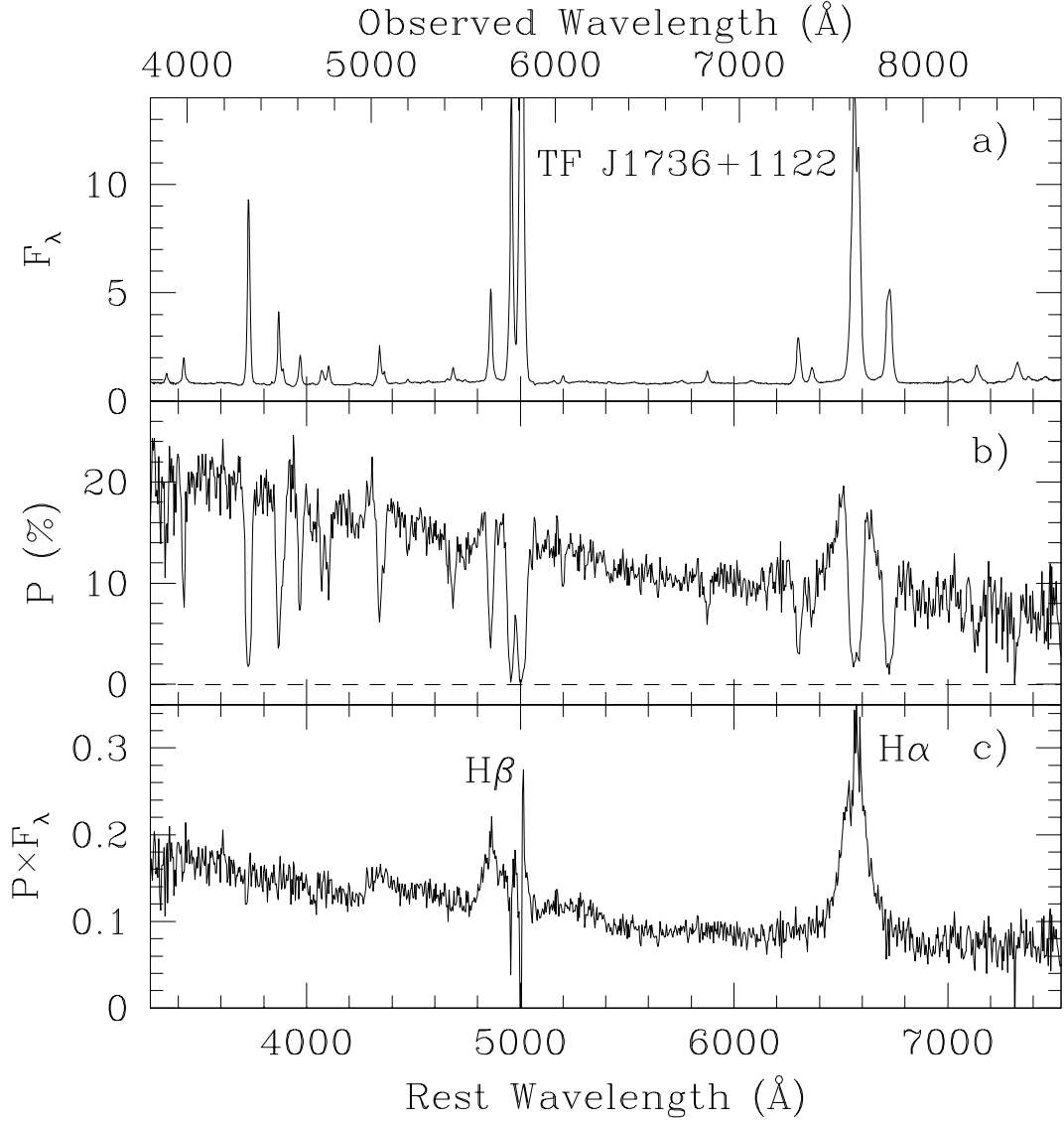}
\caption{Spectropolarimetry of TF~J1736+1122. (a) Total flux spectrum \flam,
(b) observed degree of polarization \p, and (c) polarized flux spectrum \pf.
The flux scales are in units of 10$^{-16}$ \fluxu. Fe~II emission features
are clearly present in \pf~around 4570~\AA~and 5250~\AA. \label{tf1734spol}}
\end{figure}

Unlike the two ULIRGs discussed above, TF~J1736+1122 is highly polarized, 
having a continuum \p~reaching up to $\sim$ 23\% in the near-UV. 
The spectropolarimetry is shown in Figure \ref{tf1734spol}.
The polarized flux spectrum, \pf, clearly displays not only broad 
Balmer emission lines of \ha, \hb, H$\gamma$, but also the Fe~II blends
near 4570~\AA~and 5250~\AA. In this respect, it is similar to the \pf~spectrum
of the hidden Seyfert 1 galaxy NGC 1068 (Miller, Goodrich \& Mathews 1991; 
\cite{t95}). 

We corrected the polarization for a small amount of interstellar
polarization (ISpol), based on the polarization of the narrow \oii~emission
line and the polarization of several stars near the line of sight to
TF~J1736+1122. The ISpol we adopt is \p~= 0.4 \% at PA = 80\arcdeg. 
This is consistent with the maximum polarization expected
($P_{max} < 9.0E(B-V)$; Serkowski, Mathewson, \& Ford 1975) 
from the Galactic reddening toward TF~J1736+1122 ($E(B-V)$ = 0.144; \cite{bh84}). 
Since the correction is small relative to the 
polarization of TF~J1736+1122, it does not significantly affect the results or
alter our conclusions.

We also corrected the observed polarization for the unpolarized starlight in 
the host galaxy (e.g., \cite{t95}). 
Unlike FF~J1614+3234 or TF~J1020+6436, the underlying stellar population 
of TF~J1736+1122 is dominated by old ($>$ a few Gyr) stars, as indicated by 
the lack of Balmer absorption lines. Accordingly, a selection of normal 
elliptical galaxies was tried in the fit, and the best fitting template galaxy 
is NGC 4478. We derived a galaxy fraction of 0.52 at 5500~\AA. 
Before starlight subtraction the percentage of polarization rises rapidly 
with shorter wavelength, reaching $\sim$ 23\% at 3250~\AA. This rise
in \p~is due to the decreasing dilution of starlight at shorter wavelengths, as
the starlight-corrected polarization is essentially flat. There is a rise
in polarization in the broad wings of \ha~and \hb, signifying 
the presence of a second diluting ``featureless continuum'' (FC2), 
perhaps due to hot stars (\cite{cf95}; \cite{h97}) or thermal radiation from 
hot electrons (\cite{t95}).

\begin{figure}
\plotone{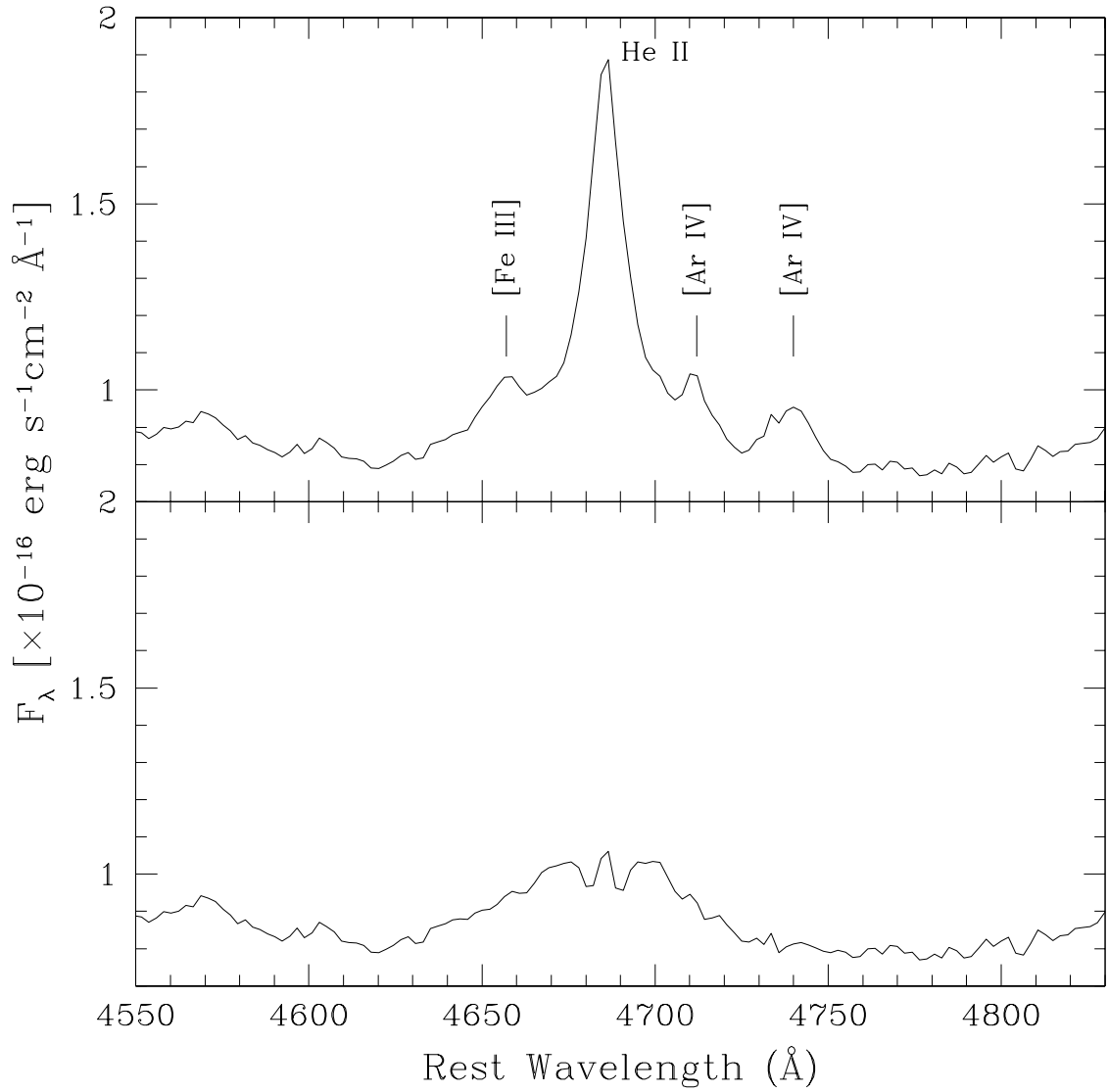}
\caption{Spectrum of TF~J1736+1122 showing the W-R bump underlying 
He~II \wave 4686. \label{wrbump}}
\end{figure}

The presence of hot stars, perhaps arising in a starburst, 
is suggested by the broad feature underlying He II \wave 4686. 
This broad feature is shown in Figure \ref{wrbump}
as a residual after the narrow line components are fitted and subtracted. 
This feature cannot be a broad scattered He~II component since it appears very 
strongly in the total flux spectrum, suggesting that the light is viewed 
directly, and it is not detected in the \pf~spectrum (Fig. \ref{tf1734spol}). 
It can be attributed to Wolf-Rayet (W-R) stars, as has been 
identified by Heckman \etal~(1997) in Mrk 477 and Storchi-Bergmann, Cid Fernandes, \& Schmitt (1998) in Mrk 1210, both of which are Seyfert 2 galaxies
with hidden broad-line regions, and for which a large fraction of FC2 
has been inferred (\cite{t95}). It seems, therefore, that there is a 
large number of young and/or W-R stars in TF~J1736+1122, which contributes 
to the observed FC2.
As discussed by Schaerer \& Vacca (1998), the presence of this feature
indicates that the time elapsed since the last burst of star formation 
cannot be more than a few Myr. 
If we assume that the broad feature is all due to the ``W-R bump'' and that
\hb~emission arises entirely from massive stellar activity
(clearly an overestimate since most of the \hb~must be due to the AGN),
we can derive an absolute lower limit to the ratio of W-R to O stars.
Using the relation derived by Schaerer \& Vacca (1998) (their Fig. 22 and 
equ. 17), the observed W-R bump/\hb~ratio of 0.14 (Table 3) gives
a W-R/O star number ratio $>$ 0.2. 
By contrast, there is little sign of the W-R bump in either 
FF~J1614+3234 or TF~J1020+6436, suggesting that the number of W-R stars 
relative to O stars must be low. 
This may indicate that few of the most massive stars ($\simgt$ 35 \msun, 
which turn into W-R stars) are formed in these galaxies, or their star 
formation episodes, which may still be ongoing, last longer than in 
TF~J1736+1122. 

\section{Discussion}
\subsection{Modelling of the Continuum and Line Polarization in TF~J1736+1122}

\begin{figure}
\plottwo{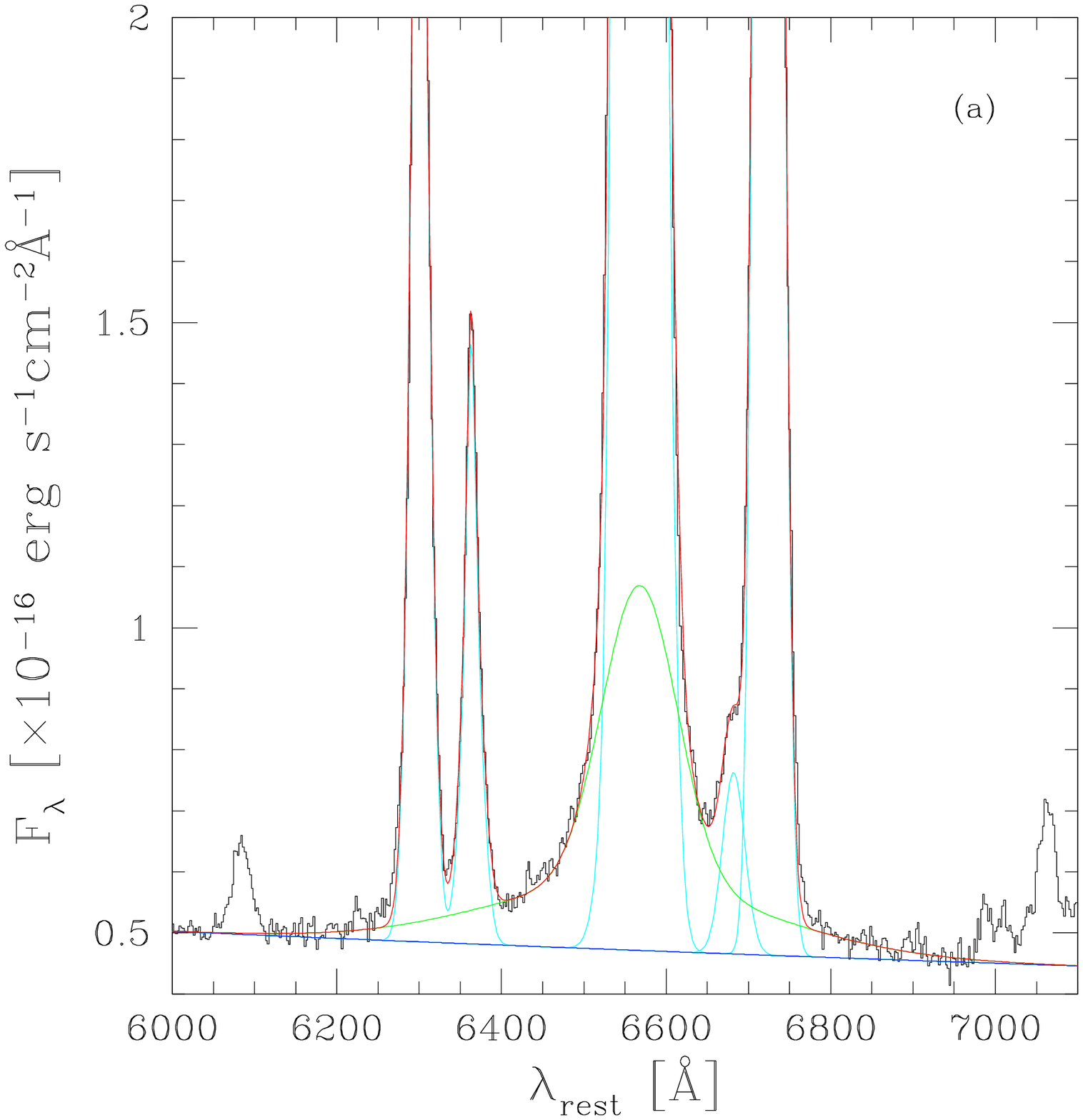}{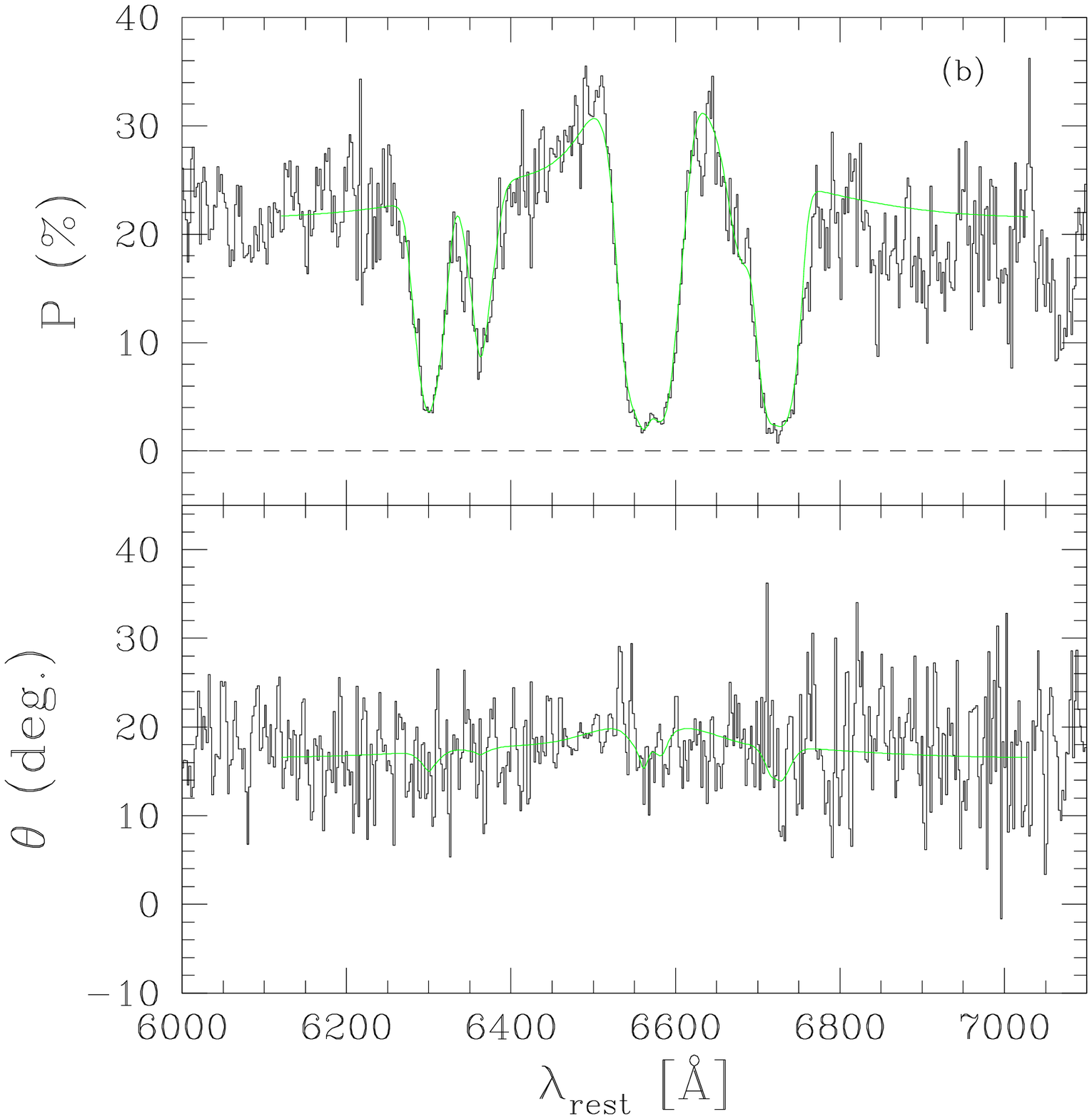}
\caption{Polarization modelling at \ha~for TF~J1736+1122.
(a) Fitting of continuum and Gaussian components to emission lines in total
flux; (b) Results of the fit (smooth curves) compared to the observed \p~\& 
\pa. The polarization has been corrected for starlight dilution. \label{polmol}}
\end{figure}

In order to separate the line and continuum components of polarization of 
TF~J1736+1122, we modelled the observed spectropolarimetric observations.  
Following a method similar to that described by Tran \etal~(1997), 
we model the spectropolarimetry surrounding the \ha~+ [N II] complex by
simultaneously fitting the continuum and emission-line components in the total 
flux, \qf, and \uf~spectra. The broad \ha~in \pf~is best fitted by two 
Gaussian profiles with FWHMs of 18000 \kms~and 4900 \kms, giving a combined
FWHM of 5580 \kms. We use this same 2-component model to fit the 
\ha~line profile in the total flux spectrum. 
The narrow lines of \ha~+ \nii, \oi, \sii~were fitted assuming 
that they are similarly polarized, and having Gaussian profiles with 
FWHMs similar to that of the \oiii~lines ($\sim$ 840 \kms). 
The results of our polarization modelling are shown in Figure \ref{polmol}.  
 
The results indicate that the broad-line polarization is 
very high, $\sim$ 50\%, while the continuum \p~is much lower at 
$\sim$ 22 \%, both at a similar PA of 20\arcdeg. 
This indicates that the amount of unpolarized FC2 flux relative to 
the total continuum (i.e., FC1 + FC2) would be about 60\% in order to make 
the continuum and broad-line polarizations equal. 
This is also just the amount required in order to scale up the polarized
flux spectrum and subtract off the broad-line component from the total flux 
spectrum. 
Our fitting also shows that the narrow lines possess a small amount of 
polarization $\sim$ 0.4 \% at PA = 162\arcdeg. 
They are clearly polarized at a PA different from that of 
the broad line and continuum, as evidenced by the apparent absorption 
in the \pf~spectrum at the wavelengths of strong 
narrow emission lines such as \oiii~\waves 4959, 5007 (Fig. \ref{tf1734spol}).
The results found here are similar to all prior results for Seyfert 2 and
narrow line radio galaxies, (e.g., \cite{t95}; Tran, Cohen, \& 
Goodrich 1995; \cite{cim96}; \cite{dey96}; \cite{t98}) in that they all show 
the presence of an FC2, an old stellar population, high polarization in the 
continuum and broad lines but little or no polarization in the narrow lines.

Polarization modelling at \hb~+ \oiii~shows similar results for 
the broad-line and continuum
polarization. However, the fitting here is complicated by the presence
of strong underlying optical Fe~II emission, and thus the results are not as 
well constrained. However, the similar polarizations obtained suggest
that the polarization is independent of wavelength.

The continuum polarization PA shows a slight rotation of 
$\sim$ 4\arcdeg~from 16\arcdeg~to 20\arcdeg~across the wavelength range 
covered. This may mean that FC2 is slightly polarized
or that there is some small interstellar polarization internal to the 
host galaxy. Any of these effects is expected to be small since there
is little or no difference in the polarization PA between the continuum and 
broad lines at \ha~or \hb. 

\subsection{Polarized Flux Spectrum and Fe~II emission in TF~J1736+1122}

The luminosity of the hidden broad-line AGN indicates that it is consistent
with being a quasar. Using the formulation of V\'eron-Cetty \& V\'eron (1996),
the observed absolute magnitude of TF~J1736+1122 is $M_B \approx -21.2$. 
Assuming an optically thin, uniformly filled
scattering cone with half-opening angle of 40\arcdeg~oriented at an 
inclination of 70\arcdeg~(\S 4.4), producing an intrinsic polarization of 50\%,
we derive $f_{direct} / f_{scattered} \approx 10$ (see e.g., \cite{bro98}).
Thus the luminosity of the obscured source is $M_B \approx -23.7$, 
marginally qualifying its status as a quasar.  
With an optical spectral index of $\alpha \sim -0.24$ 
($f_\nu \propto \nu^\alpha$), the scattered (\pf) spectrum of TF~J1736+1122 
also appears very similar to the composite
quasar spectrum of Francis \etal~(1991), which has $\alpha = -0.32$.
The Balmer decrement in \pf~is 4.2 $\pm$ 1.
These results indicate that there is no significant ``bluening'' or reddening 
of the scattered spectrum. Thus it appears that the scatterer is gray, as 
would be the case for electrons. 
For comparison, the Balmer decrement in the NLR is \ha/\hb~= 2.9, also 
essentially unreddened by dust.

As has been mentioned, an interesting feature in the \pf~spectrum of 
TF~J1736+1122 is the presence of strong polarized Fe~II emission. 
The rest-frame EWs of the Fe~II blends at 4570~\AA~and 5250~\AA~are 
$\approx$ 35~\AA~and 50~\AA, respectively\footnote{The Fe~II EWs were measured
by scaling and broadening the Fe~II template I Zw 1 to fit the observed blends,
as described by Boroson \& Green (1992).}.
This is comparable to those seen in normal quasars,
Seyfert 1, and broad-line radio galaxies (e.g., \cite{jol91}). 
The detection of Fe~II in \pf~is rather surprisingly rare among the class 
of polarized broad-line AGNs.
Of all the Seyfert 2s and narrow-line radio galaxies known to 
show HBLRs (\cite{t95}; \cite{tcg95}; \cite{y96}; \cite{coh98}) only NGC 1068 
and Mrk 463E have been observed previously to show optical Fe~II in 
their \pf~spectra. 

Although optical Fe~II emission in AGNs has been known for many years,
its exact nature and origin are still not fully understood.
The strength of Fe~II emission tends to be weaker in 
broad-line radio galaxies compared to radio-quiet Seyfert 1 galaxies 
(e.g., \cite{deo77}). 
This general tendency between Fe~II strength and radio-loudness also
holds for QSO and quasars (\cite{bg92}), but radio-loud objects are by no
means always weak Fe~II emitters (e.g., \cite{jol91}).
Among the radio-loud quasars, it has been found (e.g., \cite{h83}; \cite{jb91};
\cite{bro96}) that Fe~II strength is correlated with radio spectral index, 
in the sense that flatter-spectrum sources display stronger optical Fe~II.
TF~J1736+1122 has a radio spectrum that is rather steep (Table 2). 
Its radio-to-optical flux ratio and radio luminosity (Table 2) 
indicate that TF~J1736+1122 is formally radio-loud, albeit only marginally. 
In the orientation-dependent model of quasars, flat-spectrum radio sources are 
thought to be those viewed at small inclination angle where the core
component dominates, while steep-spectrum radio sources are those 
viewed at larger viewing angle where the lobes dominate 
(e.g., \cite{ob82}; \cite{bo84}). Thus, TF~J1736+1122 could be a radio-loud 
quasar being viewed at large inclination angle, with heavy obscuration by 
dust, perhaps in the form of a putative torus, dominating the central nucleus.

Because Fe~II emission appears prominently in broad-line (type 1) AGNs, 
in which the nuclear and BLR radiation escape freely in 
our direction, it is thought to be associated closely with the 
ionizing continuum, BLR and radio core (e.g., \cite{zk91}).  
However, several observations (\cite{hh87}; \cite{cs97}) have shown
that Fe~II emission in the QSO PG 1700+518 varies spatially, 
suggesting that it could come from 
more extended regions, perhaps in a superwind and shocks generated by 
supernovae and supernova remnants of a starburst (\cite{ter92}).
Since the polarization magnitude and position angle of the continuum, broad 
emission lines, and Fe~II emission are virtually identical in TF~J1736+1122, 
indicating that they must all follow the same scattering geometry, 
our results support the picture in which Fe~II emission originates 
from a compact source very close to the nucleus, arguing against an extended 
supernova origin for Fe~II.

\subsection{Detection of HBLRs and Nature of ULIRGs: Starburst versus AGN}

The original motivation for the HBLR search of ULIRGs in the present study 
was initiated by their positions in the radio-FIR correlation diagram. 
Based solely on their general coincidence with the 
hyperluminous $IRAS$ galaxies in this diagram, and Sanders \etal~(1988) 
ULIRG-quasar evolutionary scheme, we would expect them to 
harbor obscured quasars.
This expectation has not been borne out.
The lack of polarization in TF~J1020+6436 and FF~J1614+3234 raises an important question: 
if a BLR is not detected in polarized light, does it truly not exist?
While we cannot rule out the presence of an obscured quasar in these sources,
it appears that if a quasar BLR exists, its presence ought to 
betray itself by some tell-tale signs.
With sufficiently high signal-to-noise ratio (S/N), a BLR should be readily 
observed if it is present. 
It could be that the polarized signal is swamped by unpolarized light
and is undetectable, or perhaps other factors such as the geometry 
and distribution of the surrounding medium are less than optimal for 
scattered light to reach our line of sight. 
In such cases, little or no light from the central BLR could be observed even 
if it existed. 
There are reasons to suggest that although this situation may well be 
possible, it is unlikely. First, if the quasar ``monster'' is present, 
deep, high S/N observation at Keck should be able to reveal it, even in 
the presence of a relatively large starlight fraction. 
This has been demonstrated by the spectacular detection
of broad Balmer lines in Cygnus A (\cite{ogl97}), which have 
long been sought after in previous fruitless attempts 
(\cite{gm89}; \cite{jt93}).
At the level of S/N of the present observations, we should have 
easily detected the broad lines if they were present at a polarized flux level 
similar to that in Cygnus A. 
Second, a detection of broad polarized emission lines in an object dominated by 
a strong Balmer absorption-line (A star) spectrum would be unprecedented. 
Of the approximately two dozen HBLR type 2 AGNs detected by spectropolarimetry to date 
(e.g., \cite{t95}; \cite{tcg95}; \cite{y96}; \cite{t98}; \cite{coh98}), 
{\it none} fits this description. The most likely, and reasonable 
explanation for this is that such objects do not exist or are very rare. 
The similar galaxy fractions detected in systems both with and without 
strong Balmer absorption lines (\S 3.2, 3.3) make it unlikely that 
the lack of detectable polarization (and hence of hidden quasar) in the 
former is simply due to the large dilution of scattered light by hot stars.
If we believe in a scenario in which both the QSO and starburst activities
are triggered by close interaction and can coexist (\cite{sto98}), then 
this may imply that the life times of the ``starburst phase'' of 
QSOs are very short, comparable to those of the most massive stars. 
Third, our finding is consistent with 
the conclusions reached by the near-infrared spectroscopic study of 
Veilleux \etal~(1997), who showed that the optically LINER-like ULIRGs
do not show any evidence for broad Paschen lines or strong high-ionization 
[Si~VI] emission (which would signify the presence of a genuine monster).
By contrast, most, if not all (7--9 out of 10) of the Seyfert 
2 ULIRGs in their sample were found to display obscured BLR and strong [Si~VI].
Since TF~J1020+6436, and perhaps FF~J1614+3234 are marginal Seyfert 2s, 
borderlining LINERs, they are not expected to contain hidden BLRs.

\begin{figure}
\plotone{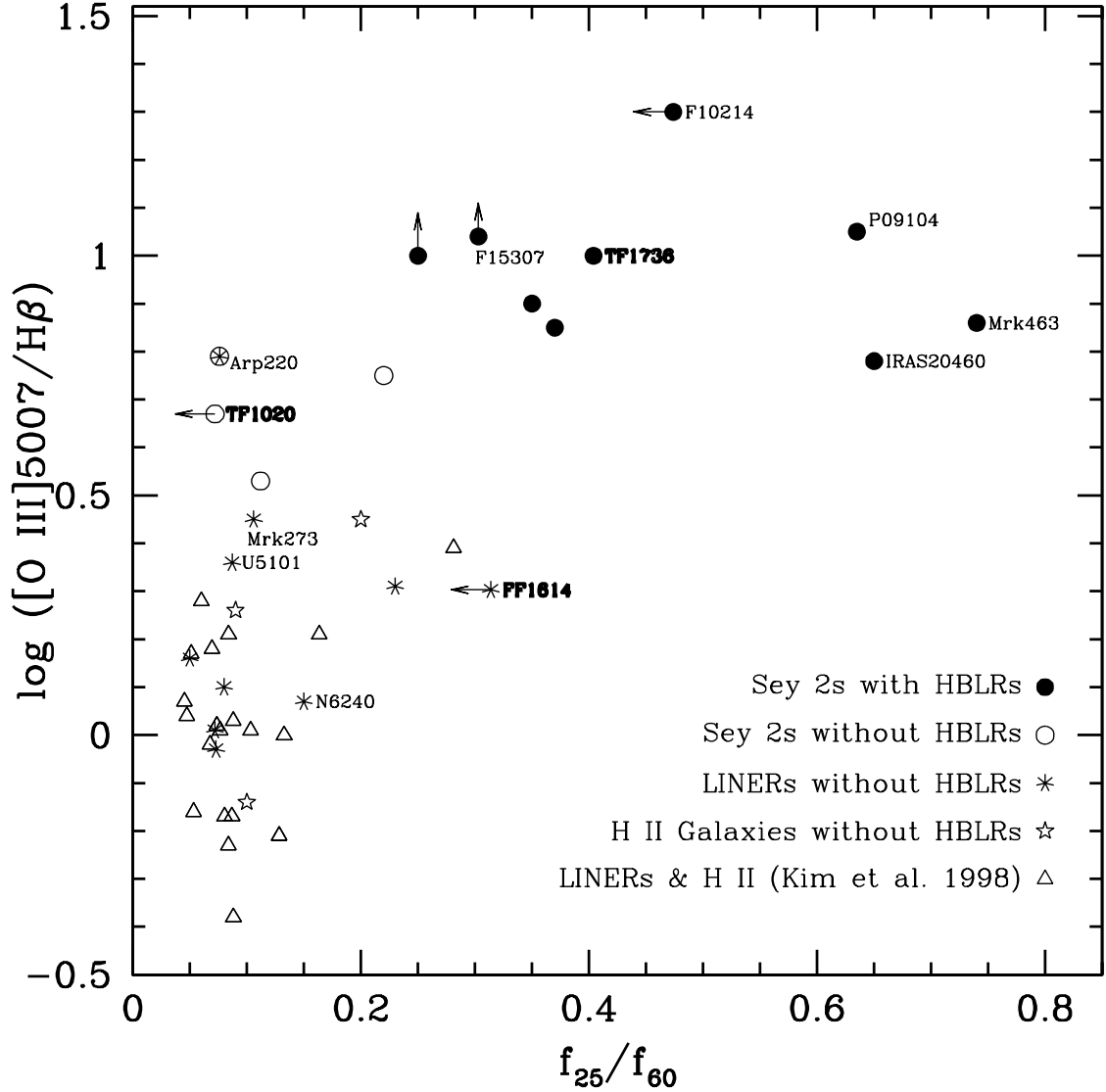}
\caption{\oiii~\wave 5007/\hb~versus IR color \firr~for narrow-line ULIRGs
in which HBLRs have been searched for. Seyfert 2s with HBLRs detected either 
with spectropolarimetry or near-IR spectroscopy are represented by solid 
circles. Those without detections of HBLRs are shown as open circles.  
Stars represent H II region galaxies and asterisks denote LINERs.
Also plotted in open triangles are all ULIRGs classified as H II and LINERs
in Kim \etal~(1998). Note the clear tendency for ULIRGs with HBLRs to have
warmer IR color and higher excitation spectrum. The opposite holds for LINERs 
and H II galaxies. \label{ionirc}}
\end{figure}

Since the ULIRG phenomenon appears to be a mix of either AGN-dominated 
and starburst-dominated events, it is of interest to see if it could be 
determined, from optical spectroscopic and $IRAS$ photometric data alone, 
whether an ULIRG harbors a genuine quasar or is powered by a starburst.
The above results suggest that the emission-line ionization level of 
an ULIRG plays an important role in identifying the main power source.
Another important factor is the IR color \firr, as 
it has been shown that the success of detecting 
HBLRs is well correlated with this ratio in both Seyfert 2 galaxies
(Heisler, Lumnsden, \& Bailey 1997) and ULIRGs (\cite{vsk97}).
Figure \ref{ionirc} shows a plot of the line ratio \oiii~\wave 5007/\hb(narrow), 
which can serve as an indicator of the ionization level, versus \firr. 
Shown are those narrow-line ULIRGs in which HBLRs have been actively searched 
for either in near-IR spectroscopy (\cite{vsk97}), or optical spectropolarimetry
(this paper). They include not only Seyfert 2 galaxies, but also sources 
classified as H II and LINERs with available line ratios and IR colors. 
Most optical spectroscopic data are obtained from Veilleux \etal~(1995) and 
Kim, Veilleux, \& Sanders (1998).
We also include a few other well-known ULIRGs (Arp 220, Mrk 273, NGC 6240, UGC 5101), the three ``hyperluminous'' sources (F10214+4724, P09104+4109, F15307+3252), 
as well as Mrk 463E, a Seyfert 2 galaxy harboring HBLR with 
$L_{IR} = 10^{11.8}$ \lsun.
As can be seen, there is a clear tendency for higher-ionization and 
``warmer'' Seyfert 2 ULIRGs to show HBLR indicative of a ``buried quasar.'' 
Furthermore, Seyfert ULIRGs without HBLRs lie in a region of the diagram 
similar to that occupied by H II and LINER galaxies, none of which has been 
found to have HBLRs. 

Figure \ref{ionirc} is analogous to the diagnostic diagram of 
Genzel \etal~(1998), who use a plot of the mid-IR line ratio 
[O~IV]~25.9~$\mu$m/[Ne II]~12.8~$\mu$m versus
the strength of the 7.7 $\mu$m PAH (polycyclic aromatic hydrocarbons) feature,
and to that of Lutz \etal~(1998), who use the same PAH feature and the
5.9$\mu$m/60$\mu$m continuum flux ratio,
to separate out ULIRGs that are predominantly powered by AGN or starburst.
Unlike these diagrams, however, Figure \ref{ionirc} not only involves the 
much more accessible (for moderate redshifts) optical lines of [O~III] and \hb,
but can also identify those AGNs harboring hidden quasars.

Thus, it appears that when an IR-luminous object displays 
a very high IR luminosity (ultraluminous), is relatively ``warm", 
and shows a high-ionization spectrum characteristic
of Seyfert 2 galaxies, it would most certainly contain an HBLR, (i.e., it is 
truly a buried quasar). TF~J1736+1122, for example,
shows a classic strong-ionization Seyfert 2 spectrum,
a warm $f_{25}/f_{60}$ color of 0.404, an IR luminosity of 
$L_{IR} \approx 10^{12}$ \lsun, and is observed to show a spectacular HBLR. 
It is highly polarized, with properties that are consistent with there 
being a quasar.
FF~J1614+3234, on the other hand, has an extremely high $L_{IR}$ ($> 10^{12.6}$ \lsun),
placing it in the company of the hyperluminous IR galaxies, and may be 
relatively warm (\firr~$<$ 0.31), but displays a feeble LINER spectrum. 
It is not polarized and there is no evidence of any hidden broad lines. 
The one characteristic that the ULIRGs with HBLR have in common is the 
display of classic high-ionization optical spectrum of a Seyfert 2. This then
appears to be a necessary condition for an ULIRG to harbor a buried quasar.

\subsection{The Influence of Quasars on Diagnostic Properties of ULIRGs} 

We now discuss how these three conditions -- ionization level, IR luminosity, 
and IR color -- are related to the existence of an obscured quasar in these 
AGNs.

a) High-ionization Seyfert 2 spectrum: 
This is necessary because a hard power-law input spectrum is needed 
to produce the high-ionization emission lines characteristic of an AGN. 
This is most likely produced by an accretion disk feeding a central black hole 
rather than by a population of hot stars, or bursts of star formation 
(with possible additional contributions from shocks).

b) High IR luminosity:
Veilleux \etal~(1995) and Kim \etal~(1998) have found that the fraction of 
Seyfert galaxies among luminous infrared galaxies increases dramatically for
$L_{IR} > 10^{12}$ \lsun. It follows naturally then that the number of 
galaxies with HBLRs also increases with IR luminosity. 
The enormous $L_{IR}$ in these objects likely arises from
reprocessed optical/UV/X-ray radiation of the ionizing continuum 
which was absorbed by the putative dusty molecular torus surrounding
the central engine (e.g., Granato, Danese, \& Franceschini 1996).

c) Warm IR color: 
Hutchings \& Neff (1991) and Kay (1994) have suggested that HBLR Seyfert
galaxies tend to show warmer IR color. This has recently been confirmed by 
Heisler \etal~(1997) in their survey of southern Seyfert 2 
galaxies. They suggest that this trend could be 
explained in the context of the obscuring torus paradigm, where Seyfert 2s
with HBLRs are those viewed at intermediate but not too high an 
inclination angle, so that our line of sight encompasses much of the inner, 
warmer regions of the dusty torus. Since the dusty torus could be optically thick 
even to mid-IR photons (12$\mu$m--25$\mu$m, \cite{dop98}), at larger viewing 
angle, our line of sight intercepts mostly the outer, cooler gas and dust in 
the torus, giving rise to a Seyfert 2 with cool IR color and no 
detectable HBLR.
This implies that the scattering takes place
very close to the central region. 
The same picture may also be true in the ULIRGs, leading to a similar
correlation between the visibility of HBLR and warmness 
(\cite{vsk97}; Fig. \ref{ionirc}).
 
The viewing inclination to these objects can in principle be
constrained by the inferred intrinsic polarization of these ULIRGs, and
thus provides a way to test the above interpretation. 
The polarizations from ULIRGs with HBLRs are generally
very high ($\sim$~20--30\%), indicating that the viewing angle is also large.
The highest intrinsic polarization observed   
is perhaps that of TF~J1736+1122 presented in this paper with \p~$\sim$~50\%.
For a uniformly filled cone of optically thin scattering material with a
typical half-opening angle of $\sim$~40\arcdeg~(e.g., \cite{wt94}),
we obtain an inclination 
angle of $i \sim$~70\arcdeg, based on the model of Brown \& McLean (1977)
(see also \cite{mg90}; \cite{mgm91}; \cite{wil92}; \cite{bk93}; \cite{hw93}; \cite{bro98}). 
This seems too high for the picture proposed by Heisler \etal~If 
multiple scattering in clumpy clouds as well as dichroic extinction
by optically thick gas contribute, the intrinsic polarization would be lower
(\cite{k95}; \cite{kis96}), and thus $i$ would have to be even higher.
Similarly, if the cone opening angle is larger than 40\arcdeg, 
the inferred $i$ would be higher for a given \p.
For smaller cone or torus opening angle, the required $i$ would be reduced,
but the fraction of scattered light intercepted by the observer would also
be much less. For example, if the half-opening angle were 20\arcdeg, $i$ would
be $\approx$~58\arcdeg, but the fraction of light scattered would be reduced 
by a factor of 3.5, making it much harder to detect the polarization.
In any case, for \p~= 50\%, $i$ cannot be smaller than $\sim$~55\arcdeg, 
at which point virtually no light gets scattered from the hidden quasar since
the cone would be infinitely narrow. 
Thus it seems likely that our viewing angle to TF~J1736+1122 is
$\simgt$~70\arcdeg, and yet broad polarized emission lines are easily seen 
and a warm IR color is detected, in contrast to the expectations of the
Heisler \etal~interpretation. 

An alternative picture is that, in the case of the ULIRGs,
the warmness of the source, as well as the high-ionization level, is a 
direct result of the very presence of a quasar ``monster" in
the nucleus, and not solely a consequence of our viewing angle.
Apparently, in order for dust grains to radiate efficiently at 25~$\mu$m,
heating sources with very high energy densities (e.g., AGN) may be required 
(cf., Rowan-Robinson \& Crawford 1989; Condon, Anderson, \& Broderick 1995).
In most ``cold" ULIRGs the quasar simply does not exist, has not yet had
sufficient time to form, or has burned out; the energetics
are instead dominated by starburst, H II region, or LINER activity.
The lack of HBLRs in these types of objects is unlikely to be due to dust 
obscuration (e.g., \cite{vsk97}), since even in such cases scattered
broad lines could still be observed.

\section{Conclusions}

Although evidence of starburst activity is found in all
three objects of the present study, some ULIRGs are powered predominantly
by quasars, and some are mainly energized by starbursts with no indication of 
an obscured AGN. 
Thus their high IR luminosity of $\sim$ 10$^{12}$--10$^{13}$~\lsun~does not 
guarantee the presence of a quasar in 
the center of these objects, such as those found in other well-known ULIRGs 
(e.g., F10214+4724, P09104+4109, and F15307+3252).
Apparently, there is a class of galaxies which have IR luminosities similar
to those of the ``classical" $IRAS$-warm galaxies but there are no detectable
quasars residing in them. The ULIRG phenomenon, therefore, can
be both explained by the presence of a quasar or a powerful starburst.

We discovered a new hidden broad-line $IRAS$ galaxy TF~J1736+1122, which is 
intrinsically highly polarized ($\sim$ 50\% in broad \ha), independent of wavelength. 
The obscured source has characteristics consistent with it being a quasar.
The detection of HBLR in this galaxy and the non-detections in FF~J1614+3234 and
TF~J1020+6436 are consistent with the idea that buried quasars preferentially reside 
in ULIRGs having warm colors, but with the necessary condition that their 
spectra must show characteristics of a high-ionization Seyfert 2 galaxy. 
Quasar-hosting galaxies also seem to have a preference for a {\it dominant} 
stellar population that is old ($>$ 1 Gyr), while galaxies without detectable 
sign of a quasar display a younger stellar population. 
If the AGN and starburst activities in these objects 
are triggered by the same mechanism at the same time (cf. \cite{sto98}), 
then our findings indicate that quasars are likely to be found only in 
evolved systems with age $\simgt$ 300 Myr.
More observations of a larger sample of ULIRGs, and especially of quasar host 
galaxies are needed to further examine this issue.

\acknowledgments

We thank Carlos ``the mad Belgian" De Breuck for fruitful discussions, and
Jan Tweed for her assistance with Figure 2.
The W. M. Keck Observatory is operated as a scientific partnership between
the California Institute of Technology and the University of California, 
made possible by the generous financial support of the W. M. Keck Foundation. 
Work performed at the Lawrence Livermore National Laboratory is supported
by the DOE under contract W7405-ENG-48.
R.A. acknowledges the support of NSF grant AST-9617160.
This research has made use of the NASA/IPAC Extragalactic
Database (NED) which is operated by the Jet Propulsion Laboratory,
California Institute of Technology, under contract with the National
Aeronautics and Space Administration.

\clearpage
\begin{deluxetable}{lcccccccc}
\tablewidth{0pc}
\tablecaption{Infrared Properties}
\tablehead{
\colhead{Object} & \colhead{$z$} & \colhead{$m^a$} & \colhead{log($L_{IR}$/\lsun)$^b$} & \colhead{$f_{12}$} & \colhead{$f_{25}$} & \colhead{$f_{60}$} & \colhead{$f_{100}$} & \colhead{$f_{25}/f_{60}$} }
\startdata
FF~J1614+3234 & 0.710 & 19.1 & 12.6 -- 13.2 & $<$ 0.065 & $<$ 0.055 & 0.174 & $<$ 0.54 & $<$ 0.31 \nl
TF~J1020+6436 & 0.153 & 19.0 & 11.9 -- 12.1 & $<$ 0.095 & $<$ 0.062 & 0.86 & 1.24 & $<$ 0.072 \nl
TF~J1736+1122 & 0.162 & 18.0 & 11.8 -- 12.3 & $<$ 0.081 & 0.196    & 0.484 & $<$ 3.31 & 0.404 \nl

\tablenotetext{}{$f_{12}$, $f_{25}$, $f_{60}$, and $f_{100}$ are the $IRAS$ 
flux densities in units of Jy at 12 $\mu$m, 25 $\mu$m, 60 $\mu$m, and 100 $\mu$m (from NASA/IPAC Extragalactic Database).}
\tablenotetext{a}{Apparent photographic magnitudes from Condon et al. (1995), 
except for FF~J1614+3234, where an $r_s$ magnitude from Stanford et al. (1998) 
is listed.}
\tablenotetext{b}{$L_{IR}$ $\equiv$ $L$(8--1000 $\mu$m) is computed from 
the observed fluxes in four $IRAS$ bands, according to the formulae given 
in Sanders \& Mirabel (1996). The lower value denotes the luminosity that 
results from setting to zero all bands with upper flux limits, and the upper
value is obtained from assuming that these limits are actual detections.}

\enddata
\end{deluxetable}

\clearpage

\begin{deluxetable}{lccccccccc}
\scriptsize
\tablewidth{0pc}
\tablecaption{Radio Properties}
\tablehead{
\colhead{Object} & \colhead{RA (J2000)} & \colhead{Dec (J2000)} & \colhead{$S_{365}^a$} & \colhead{$S_{1400}$} & \colhead{$S_{4850}$} & \colhead{$-\alpha^{365}_{1400}$} & \colhead{$-\alpha^{1400}_{4850}$} & \colhead{log $R^b$} & \colhead{log $L_{1400}^c$} }
\startdata
FF~J1614+3234 & 16 14 22.11 & +32 34 03.7 & \nodata & 1.19 $\pm$ 0.14$^d$ & \nodata & \nodata & \nodata & \nodata & 31.3 \nl
TF~J1020+6436$^e$ & 10 20 41.05 & +64 36 05.5 & 1037 $\pm$ 40 & 287 $\pm$ ~9$^g$ & 81.1 $\pm$ 0.2 & 0.96 & 1.02 & 2.70 & 32.2 \nl
TF~J1736+1122$^f$ & 17 36 54.92 & +11 22 40.0 & ~333 $\pm$ 29 & 122 $\pm$ ~4$^g$ & 35.5 $\pm$ 0.2 & 0.75 & 0.99 & 2.34 & 31.9 \nl

\tablenotetext{}{Radio flux densities $S_\nu$ are in units of mJy; spectral 
index $\alpha$ is defined for $S_\nu \propto \nu^\alpha$.}
\tablenotetext{a}{Texas Sky Survey (Douglas et al. 1996).}
\tablenotetext{b}{Ratio of 5 GHz radio to optical $B$ band flux, computed using 
eqs. (1)-(3) of Stocke et al. (1992). The usual dividing line between 
radio-loud and radio-quiet quasars is log $R \approx$ 1.}
\tablenotetext{c}{1.4 GHz radio luminosity in units of ergs s$^{-1}$ Hz$^{-1}$.
The usual dividing line between radio-loud and radio-quiet quasars is log $L_{1400}$ = 32.5.}
\tablenotetext{d}{FIRST Survey (Becker et al.~1995).}
\tablenotetext{e}{WENSS flux is 1075 $\pm$ 4 mJy at 327 MHz.}
\tablenotetext{f}{VLA 8440 MHz flux is 16.6 $\pm$ 1.0 mJy (this paper), which implies $\alpha^{4850}_{8440} = -1.45$. Thus the radio spectrum steepens rapidly between 365 MHz and 8440 MHz.}
\tablenotetext{g}{NVSS (NRAO VLA Sky Survey; Condon et al. 1998).}

\enddata
\end{deluxetable}

\clearpage

\begin{deluxetable}{llllllllllll}
\tiny
\tablecolumns{12}
\tablewidth{0pc}
\tablecaption{Emission Line Flux Ratios, Equivalent Widths and FWHMs}
\tablehead{
\colhead{} & \multicolumn{3}{c}{TF~J1736+1122$^a$} & \multicolumn{1}{c}{} & \multicolumn{3} {c}{TF~J1020+6436$^a$} & \multicolumn{1}{c}{} & \multicolumn{3}{c}{FF~J1614+3234$^a$} \\
\cline{2-4} \cline{6-8} \cline{10-12} \\
\colhead{Line} &
\colhead{Flux Ratio$^b$} & \colhead{EW$^c$} & \colhead{FWHM$^d$} & \colhead{} &
\colhead{Flux Ratio$^{b, e}$} & \colhead{EW$^c$} & \colhead{FWHM$^d$} & \colhead{} &
\colhead{Flux Ratio$^b$} & \colhead{EW$^c$} & \colhead{FWHM$^d$} 
}
\startdata
C II] \wave 2326 & \nodata & \nodata & \nodata && \nodata & \nodata & \nodata &&0.71 & 6.2 & 1390 \nl
[Ne V] \wave 3346 &  0.117    & 8.37 &  1250 && & & && & & \nl
[Ne V] \wave 3426 + O III \wave 3444 & 0.319 & 23.3  & 1040 && & & && & & \nl
[O II] \wave 3727 &  2.202 & 209 & 1080 && 2.12 (6.52)& 120 & 620 && 3.52 & 57 & 990 \nl
H11 \wave 3771    &  0.00721   & 0.67  & \nodata && & & && & & \nl
H10 \wave 3798    &  0.0152    & 1.47  &  510  &&   & & && & & \nl
H9 \wave 3835     &  0.0201    & 1.81  &  105  && & & && & & \nl  
[Ne III] \wave 3869 + He I \wave 3868  &  0.699 & 68.0 & 895 && 0.46 (1.3)& 20 & 480 && 0.62: & 10: & \nodata \nl
H8 + He I \wave 3889 &  0.175  & 17.3  &  975  && & & && & & \nl 
H$\epsilon$ \wave 3970 + [Ne III] \wave 3967 & 0.328 & 35.0 & 950 && 0.22: (0.55:) & 11: & \nodata && & & \nl 
[S II] \wave 4071 &  0.195      & 22.0 &  1280 && 0.14: (0.31:) & 6.6: & \nodata && & & \nl
H$\delta$ \wave 4102 &  0.192   & 21.8 &  920  && 0.22: (0.48:) & 9.8: & \nodata && & & \nl 
[Fe V] \wave 4227    &  0.0193  & 2.20 &  570  && & & && & & \nl 
H$\gamma$  \wave 4340 &  0.389  & 47.1 &  870  && 0.36 (0.62) & 17 & 530 && 0.78: & 15: & \nodata \nl 
[O III] \wave 4363    &  0.123  & 14.8 &  940  && 0.19: (0.32:) & 9: & \nodata && & & \nl 
He I \wave 4471     &  0.0362   & 4.75 &  930  && & & && & & \nl 
[Fe III] \wave 4658 &  0.00999  & 1.23 &  420  && & & && & & \nl 
He II \wave 4686n   &  0.0961  & 10.9 & 525 && 0.13 (0.16)& 5.0 & 240 && & & \nl
WR bump             &  0.140   & 20.4 &  3990 && & & && & & \nl 
[Ar IV] \wave 4712  &  0.00804   & 1.00 & \nodata && & & && & & \nl
[Ar IV] \wave 4740  &  0.0190    & 2.70 &   710  &&  & & && & & \nl
H$\beta$(n) \wave 4861 & 1.00 & 140 & 870 && 1.00 (1.00)& 36.0 & 340 && 1.00 & 21: & 780 \nl
[O III] \wave 4959  &  3.357     & 600.0 &  840  && 1.76 (1.61)& 63.0 & 430 && 0.93 & 20.0 & \nodata \nl
[O III] \wave 5007  &  9.886     & 1750 &  820  &&  5.37 (4.71)& 186 & 420 && 2.01 & 45.0 & 860 \nl 
[Fe VII] \wave 5159 &  0.0659    & 11.2  &  2160 &&  & & && & & \nl
[N I] \wave 5200    &  0.0784    & 13.4  &  790  &&  0.35 (0.26)& 11 & 450 && & & \nl
[Fe III] \wave 5270 &  0.0239    & 4.2   &  675  && & & && & & \nl 
He II  \wave 5412   &  0.0158    & 3.1   &  600  && & & && & & \nl 
[Cl III] \waves 5518, 5538 & 0.0248 &  5.17 & \nodata && & & && & & \nl
[Fe VII] \wave 5721 & 0.0240    & 5.20  &  1420 && & & && & & \nl
[N II] \wave 5755   & 0.0358    & 7.80  &  1000  && 0.13 (0.067) & 2.7 & \nodata && & & \nl
He I \wave 5876     & 0.177     & 38.6  & \nodata && 0.22 (0.11)& 4.5 & 150 && & & \nl
He II \wave 6074 + [Fe VII] \wave 6087 & 0.0401  & 9.31 & 1270 && & & && & & \nl
[O I] \wave 6300  & 0.579  & 123.5 &  880 &&  1.74 (0.70)& 32.7 & 260 && & & \nl
[O I] \wave 6364  & 0.199  & 42.5  &  895 &&  0.57 (0.22)& 10.6 & 260 && & & \nl
[N II] \wave 6548 & 0.951  & 230   &  1090 && 3.60 (1.30)& 70.0 & 400 && & & \nl
H$\alpha$(n) \wave 6563 & 2.89  & 700.0 &  725  && 8.77 (3.10)& 171 & 240 && & & \nl
H$\alpha$(b) \wave 6563 & 0.882 & 214   &  5580 && & & && & & \nl
[N II] \wave 6583   & 2.607     & 630  &  1170 && 10.8 (3.84)& 211 & 450 && & & \nl
He I \wave 6678     & 0.0888    & 21.8  &  1330 && & & && & & \nl 
[S II] \wave 6716   & 0.469     & 120  &  740  &&  3.09 (1.02)& 47.6 & 250 && & & \nl
[S II] \wave 6731   & 1.161     & 280  &  1010 &&  2.68 (0.88)& 41.1 & 250 && & & \nl
[Ar V] \wave 7006?  & 0.0236    & 5.8   & \nodata && & & && & & \nl
He I \wave 7065     & 0.0515    & 11.9  &  865  && & & && & & \nl
[Ar III] \wave 7136 & 0.277     & 62.0  &  1130 && 0.28 (0.076)& 4.1 & \nodata&& & & \nl
[O II] \waves 7320, 7330 & 0.315 & 64.0 &  1140 && 0.99 (0.26)& 13 & \nodata  && & & \nl
[Ni II] \wave 7378  & 0.0372    & 7.4   &  550  && & & && & & \nl
[Fe II] \wave 7453  & 0.0601    & 12.2  &  1090 && & & && & & \nl 

\tablenotetext{a}{Flux ratios and EWs are measured from starlight-subtracted spectrum.}
\tablenotetext{b}{Observed line flux ratios relative to H$\beta$; colon denotes uncertainty $\simgt$ 20\%.}
\tablenotetext{c}{Rest-frame equivalent widths in \AA.}
\tablenotetext{d}{FWHMs (\kms) have been corrected for instrumental resolution of 550 \kms.}
\tablenotetext{e}{Numbers in parentheses denote values after correction for
reddening $E(B-V) = 1.05$, based on the observed Balmer decrement \ha/\hb = 8.77.}

\enddata

\end{deluxetable}

\clearpage

\end{document}